\documentclass[twocolumn]{aastex62}
\usepackage{amsmath,amstext}
\usepackage{units}
\usepackage{url}
\usepackage{xspace}
\usepackage{graphicx}

\newcommand{\E}[1]{\times10^{#1}}
\newcommand{\msol}{ \, M_\odot}

\newcommand{\bi}{\begin{itemize}}
\newcommand{\ei}{\end{itemize}}
\newcommand{\commentOut}[1]{}

\newcommand{\flash}{\texttt{FLASH}\xspace}

\newcommand{\mesa}{\texttt{MESA}\xspace}

\shortauthors{Shen et al.}

\begin{document}

% -----------------------------------------------------------
% -----------------------------------------------------------

\title{\bf \Large{Almost All Carbon/Oxygen White Dwarfs Can Host Double Detonations}}

\correspondingauthor{Ken~J.~Shen}
\email{kenshen@astro.berkeley.edu}

\author[0000-0002-9632-6106]{Ken~J.~Shen}
\affiliation{Department of Astronomy and Theoretical Astrophysics Center, University of California, Berkeley, CA 94720, USA}

\author[0000-0002-1184-0692]{Samuel J.\ Boos}
\affiliation{Department of Physics \& Astronomy, University of Alabama, Tuscaloosa, AL, USA}

\author[0000-0002-9538-5948]{Dean M.\ Townsley}
\affiliation{Department of Physics \& Astronomy, University of Alabama, Tuscaloosa, AL, USA}

% -----------------------------------------------------------
% -----------------------------------------------------------

\begin{abstract}

Double detonations of sub-Chandrasekhar-mass white dwarfs (WDs) in unstably mass-transferring double  WD binaries have become one of the leading contenders to explain most Type Ia supernovae.  However, past theoretical studies of the explosion process have assumed relatively ad hoc initial conditions for the helium shells in which the double detonations begin.  In this work, we construct realistic C/O WDs to use as the starting points for multidimensional double detonation simulations.  We supplement these with simplified one-dimensional detonation calculations to gain a physical understanding of the conditions under which shell detonations can propagate successfully.  We find that C/O WDs $\lesssim 1.0 \msol$, which make up the majority of C/O WDs, are born with structures that can support double detonations.  More massive C/O WDs require $\sim 10^{-3} \msol $ of accretion before detonations can successfully propagate in their shells, but such accretion may be common in the double WD binaries that host massive WDs.  Our findings strongly suggest that if the direct impact accretion stream reaches high enough temperatures and densities during mass transfer from one WD to another, the accreting WD will undergo a double detonation.  Furthermore, if the companion is also a C/O WD $\lesssim 1.0 \msol$, it will undergo its own double detonation when impacted by the ejecta from the first explosion.  Exceptions to this outcome may explain the newly discovered class of hypervelocity supernova survivors.

\end{abstract}

% -----------------------------------------------------------
% -----------------------------------------------------------

\section{Introduction}

The identity of Type Ia supernova (SN~Ia) progenitors remains an unsolved mystery (see, e.g., \citealt{liu23b} for a recent review), but recent work appears to be converging towards a solution.  Many theoretical SN~Ia explosion mechanisms fall into two categories: deflagrations and/or detonations ignited close to the center of near-Chandrasekhar-mass white dwarfs (WDs; e.g., \citealt{wi73,it84,nty84,webb84}) and double detonations of sub-Chandrasekhar-mass WDs (e.g., \citealt{nomo82b,wtw86,livn90}), in which a helium shell detonation triggers a subsequent carbon/oxygen core detonation.  Double detonations were initially envisioned as occurring due to helium-rich accretion from non-degenerate donors.  However, such systems lead to relatively massive helium shells $\sim 0.1 \msol$, and the ashes of these thick shell detonations result in discrepancies when compared to observations of SNe~Ia \citep{hk96,nuge97}.  Recent theoretical work has explored the alternative possibility of degenerate donors leading to double detonations, either via stable \citep{bild07,fhr07,sb09b,shen10,fink10} or unstable \citep{guil10,dan11,dan12,rask12,pakm13a,moll14a,tani18b,tani19a,pakm21a,pakm22a} mass transfer.  The helium shells at the time of explosion in these systems are much less massive, and thus better matches to observations can be achieved.

There is now mounting evidence that the core spectral subclasses of SNe~Ia, from subluminous SN~1991bg-likes to overluminous SN~1991T-likes and SN~2003fg-likes,\footnote{Recent discoveries have uncovered classes of transients that share some spectroscopic similarities to the core spectral subgroups of SNe~Ia, including SNe~Iax or SN~2002cx-likes, SN~2002es-likes, and calcium-strong transients.  While these explosions are likely thermonuclear in origin, it is widely believed that they arise from different progenitors than the main subclasses of SNe~Ia discussed here (e.g., \citealt{taub17a}).}  are the results of detonations of sub-Chandrasekhar-mass WDs in double WD systems.  The growing support includes the lack of bright non-degenerate surviving companions within SN remnants \citep{sp12,kerz12a,kerz13a,kerz14c}, the prediction and discovery of a growing number of hypervelocity surviving companion WDs \citep{shen18b,elba23a}, and explosion simulations and radiative transfer calculations that broadly match observed features of SNe Ia, with calculations that include more accurate physics better reproducing observables \citep{fhr07,fink10,guil10,krom10,sim10,tmb12,moor13a,pakm13a,shen14b,shen18a,tani18b,poli19a,tani19a,town19a,gron20a,gron21a,shen21a,shen21b,boos21a,pakm22a,roy22a,burm23a,boos24a}.

Past studies that have included helium shells have assumed ad hoc masses and/or compositions, often appealing to a previous phase of accretion before the explosion.  However, in the case of double WD mergers, the amount of mass transferred in the lead-up to the exponential increase in accretion rate is minimal, and furthermore, the mass that is transferred passes through an accretion shock and is thus much hotter and less dense than assumed.  The true masses of the helium layers in double WD binaries at the time of explosion should in fact be much smaller than assumed in these studies, by orders of magnitude in some cases.

In this paper, we improve upon previous work by constructing realistic C/O WDs and using these as the starting points for two-dimensional double detonation calculations.  We find that C/O WDs $\lesssim 1.0 \msol$, which make up the majority of C/O WDs, are born with composition and density profiles that support helium shell detonations without the need for any added mass, and that these shell detonations robustly trigger core detonations.  WDs that are more massive than $\sim 1.0 \msol$ require the addition of $\sim 10^{-3} \msol$ of material before they can support a shell detonation, but such accretion prior to the formation of the double WD binary may be fairly common for these masses.

In Section \ref{sec:mesa}, we describe the stellar evolution calculations that produce the C/O WDs we use as initial conditions for one-dimensional detonation calculations, described in Section \ref{sec:1D}, and multidimensional detonation simulations, detailed in Section \ref{sec:2D}.  We discuss implications of our results in Section \ref{sec:disc} and conclude in Section~\ref{sec:conc}.

% -----------------------------------------------------------
% -----------------------------------------------------------

\section{Constructing C/O white dwarfs with realistic composition profiles}
\label{sec:mesa}

To set the stage for the detonation calculations that we discuss in later sections, we first create C/O WD models with realistic composition profiles using the stellar evolution code \mesa \citep{paxt11,paxt13,paxt15a,paxt18a,paxt19a}.\footnote{https://docs.mesastar.org, version 15140.}  We follow a similar recipe as in \cite{shen23a} to construct our models; see their appendix for sample inlists.\footnote{See also the materials associated with Shen's 2015 \mesa summer school lectures: https://doi.org/10.5281/zenodo.2603640.}  We begin with stars of pure $^4$He and $^{14}$N with mass fractions of $X_{\rm 4He} = 0.99$ and $X_{\rm 14N} = 0.01$, as appropriate for stars that have undergone CNO-burning, with masses of 0.5, 0.6, 0.7, 0.8, 0.9, and $1.0 \msol$.  Semiconvection, thermohaline mixing, and diffusion are turned off during this first stage, and the \texttt{mesa\_45.net} nuclear network is used.  The models are evolved through the core- and shell-burning phases until the remaining helium-burning shell is too small to support a stable steady-state solution.  This minimum-mass stably burning helium shell is a robust feature and only depends on the underlying core mass \citep{sb07,kipp13a}, thus justifying our initial conditions, which skip over the intricacies of binary stellar evolution.

After shell-burning quenches, the stars contract and reach a maximum effective temperature and then begin to evolve down the WD cooling track.  At the time of maximum effective temperature, we turn on semiconvection and thermohaline mixing along with diffusion.  Finally, we stop the models after they reach 1\,Gyr of age.

We also consider a C/O WD model of $1.1 \msol$.  If created during single stellar evolution, such a massive core will ignite carbon-burning on the asymptotic giant branch  and thus become an O/Ne WD (e.g., \citealt{mura68a,dohe15a}).  However, a C/O WD of this mass can be created if $\sim 0.1 \msol$ of helium is accreted onto a lower-mass C/O WD and is burned to C/O stably \citep{wu22a,shen23a} without igniting carbon-burning.  Thus, we construct a $1.1 \msol$ C/O WD by accreting helium-rich material ($X_{\rm 4He} = 0.99$ and $X_{\rm 14N} = 0.01$) onto our $1.0 \msol$ C/O WD model at a rate of $2\E{-6} \msol \, {\rm yr}^{-1}$, which yields a stable-burning envelope \citep{sb07,pier14a,broo16a}.  Mixing and diffusion are turned off during this phase.  Once the total mass reaches $1.1 \msol$, the star is allowed to cool down (with mixing and diffusion reactivated at the time of maximum effective temperature) and become a WD.  The model only cools for $6\E{8}$\,yr before convergence issues halt further evolution, but we deem this age to be adequate for our purposes.

\begin{figure}
  \centering
  \includegraphics[width=\columnwidth]{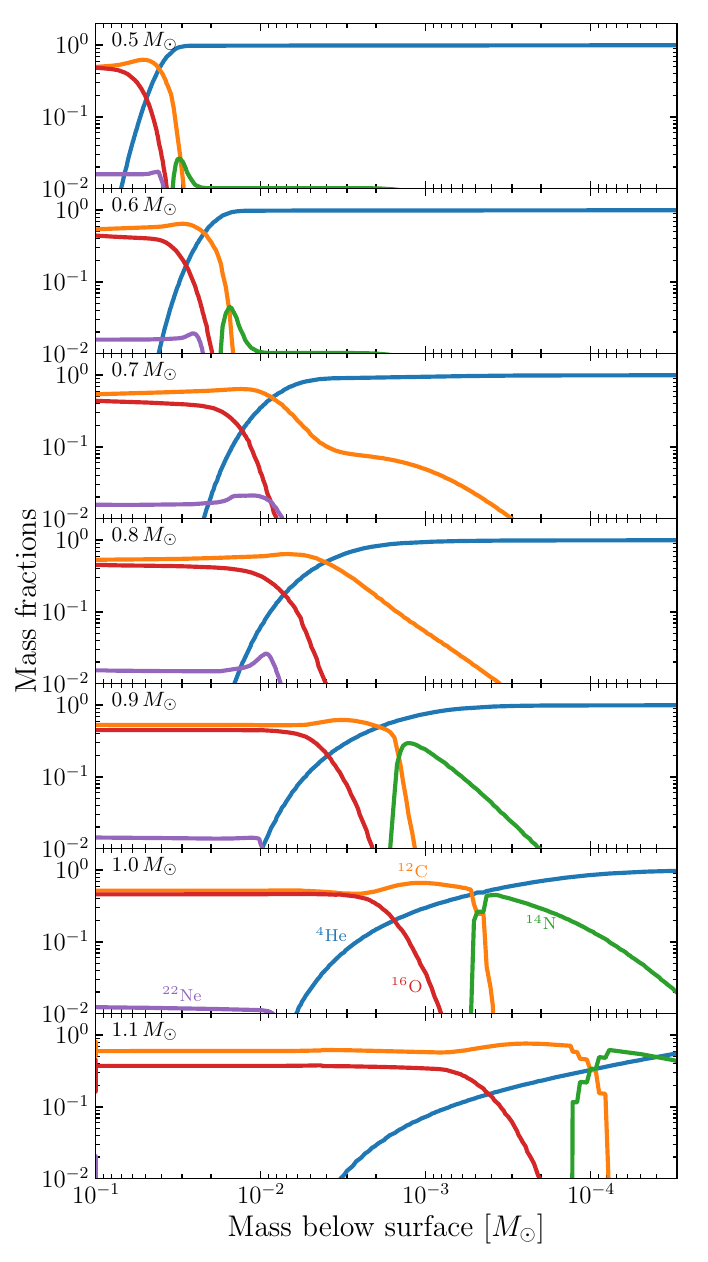}
	\caption{Composition profiles vs.\ mass below the surface of our WD models with total masses given in the upper left corner of each panel.  Mass fractions are as labeled in the $1.0 \msol$ panel.}
	\label{fig:abunvsm}
\end{figure}

\begin{figure}
  \centering
  \includegraphics[width=\columnwidth]{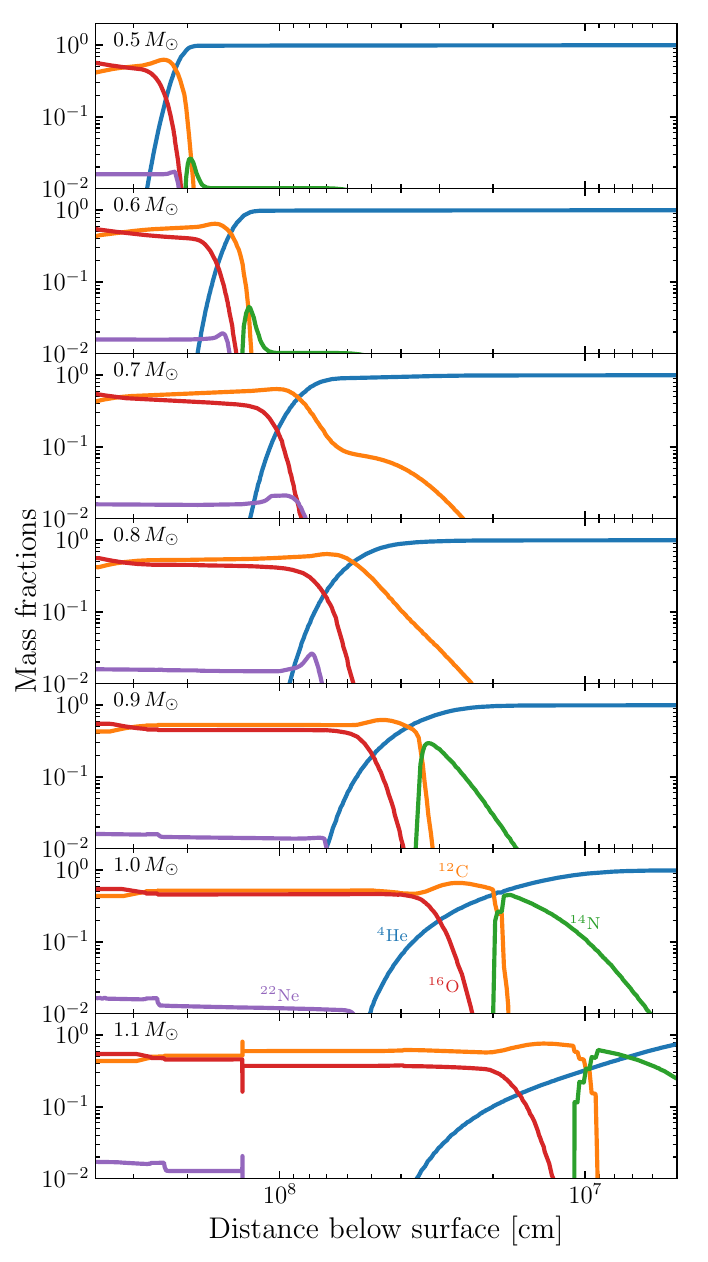}
	\caption{Same as Fig.\ \ref{fig:abunvsm} but vs.\ distance below the surface.}
	\label{fig:abunvsr}
\end{figure}

Figures \ref{fig:abunvsm} and \ref{fig:abunvsr} show mass fractions for our seven WD models as a function of mass and distance below the surface, respectively.  The helium-rich envelope is a significant fraction of lower-mass WDs, making up almost 10\% of a $0.5 \msol$ WD's mass.  The helium fraction decreases as the WD mass increases and makes up only $\sim 10^{-4}$ of a $1.1 \msol$ WD (although the tail of the helium distribution reaches much farther into the core).

\begin{figure}
  \centering
  \includegraphics[width=\columnwidth]{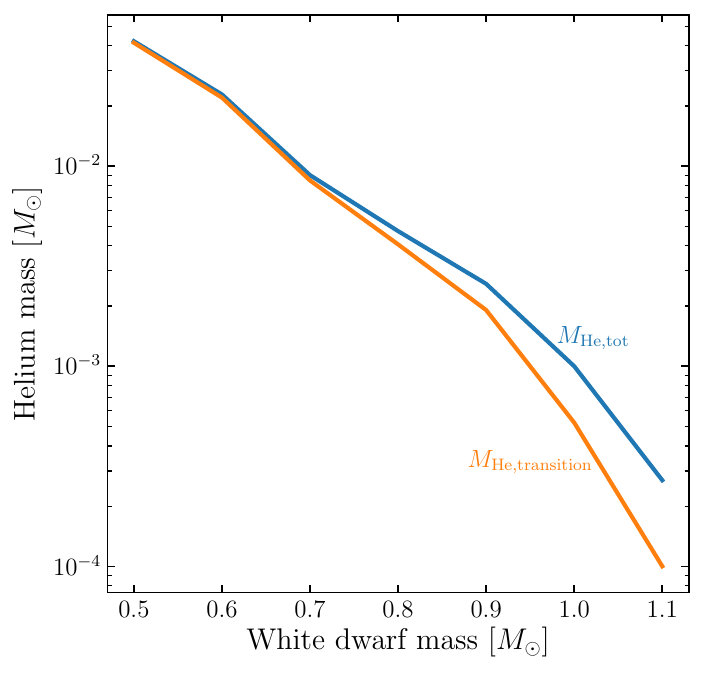}
	\caption{Total mass of $^4$He ($M_{\rm He,tot}$) and mass above the transition layer ($M_{\rm He,transition}$) vs.\ WD mass.}
	\label{fig:mhevsmwd}
\end{figure}

Figure \ref{fig:mhevsmwd} shows the total $^4$He mass, $M_{\rm He,tot}$, and the mass above the radius where the $^4$He mass fraction equals that of $^{12}$C, $M_{\rm He,transition}$, vs.\ WD mass.  The two quantities are quite close in value for most WDs, but as the WD mass increases, the total helium mass becomes several times larger than the mass above the transition radius.  This is due to the aforementioned significant helium tail that extends into the WD core.

\cite{zena19b} also emphasized the importance of the surface helium layers on C/O WDs by constructing  models  with \mesa.  Within our region of overlap ($0.5-0.7 \msol$), our higher-mass models are in rough agreement, but we obtain significantly smaller helium masses at the lower-mass end; e.g., our $0.5 \msol$ model has 50\% less helium than obtained by \cite{zena19b}.  The reasons for this discrepancy are unclear.

With our realistic C/O WD models in hand, we are ready to construct two-dimensional detonation simulations, which we perform in Section \ref{sec:2D}.  But first, we build our physical understanding with simplified one-dimensional detonation calculations, which we now describe.

% -----------------------------------------------------------
% -----------------------------------------------------------

\section{One-dimensional He/C detonations}
\label{sec:1D}

In this section, we examine the propagation of one-dimensional detonations in material with equal mass fractions of $^4$He and $^{12}$C.  This choice is motivated by the two-dimensional detonations discussed in the next section, which propagate in the transition region between the helium-rich surface layer and the carbon-rich core.  Here, we calculate whether or not detonations can propagate at various densities, their associated lengthscales, and the effects of resolution on these results.

Our one-dimensional detonation simulations are performed using the reactive hydrodynamics code \flash \citep{fryx00,dube09a},\footnote{https://flash.rochester.edu/site/flashcode, version 4.3.} modified to use nuclear networks and  tabulated  reaction rates from \mesa \citep{shen18a,town19a,boos21a,boos24a}.  We use a 56-isotope nuclear network for all of the simulations discussed in this paper,  composed of neutrons, $^1$H, $^4$He, $^{11}$B, $^{12-13}$C, $^{13-15}$N, $^{15-17}$O, $^{18-19}$F, $^{19-22}$Ne, $^{22-23}$Na, $^{23-26}$Mg, $^{25-27}$Al, $^{28-30}$Si, $^{29-31}$P, $^{31-33}$S, $^{33-35}$Cl, $^{36-39}$Ar, $^{39}$K, $^{40}$Ca, $^{43}$Sc, $^{44}$Ti, $^{47}$V, $^{48}$Cr, $^{51}$Mn, $^{52,56}$Fe, $^{55}$Co, and $^{56,58-59}$Ni; this network is the same as that used in \cite{town19a} and \cite{boos21a,boos24a}, with the addition of the isotope $^{19}$F for completeness, although it does not appear to have an effect on our results.  Detonations are initiated in spherical symmetry via a hotspot in a uniform medium with a temperature of $  \unit[2\E{9}]{\rm K}$ at the center that decreases linearly to the edge of the hotspot to a value of  $\unit[10^6]{\rm K}$, matching the temperature of the unperturbed material.  The size of the hotspot is chosen to be approximately the smallest size that results in a successful detonation that continues to propagate for at least 10 hotspot radii or 100 sonic lengthscales, as defined below.  Burning is turned off within the shock, and no reaction limiter is used \citep{kush13a,shen18a,kush20b,boos21a,boos24a}, although both choices are moot for our highest resolution cases because the burning structure is resolved.  Electron screening is implemented following \cite{chug07a}, and neutrino losses are neglected given the high temperatures achieved in the detonations.

\begin{figure}
  \centering
  \includegraphics[width=\columnwidth]{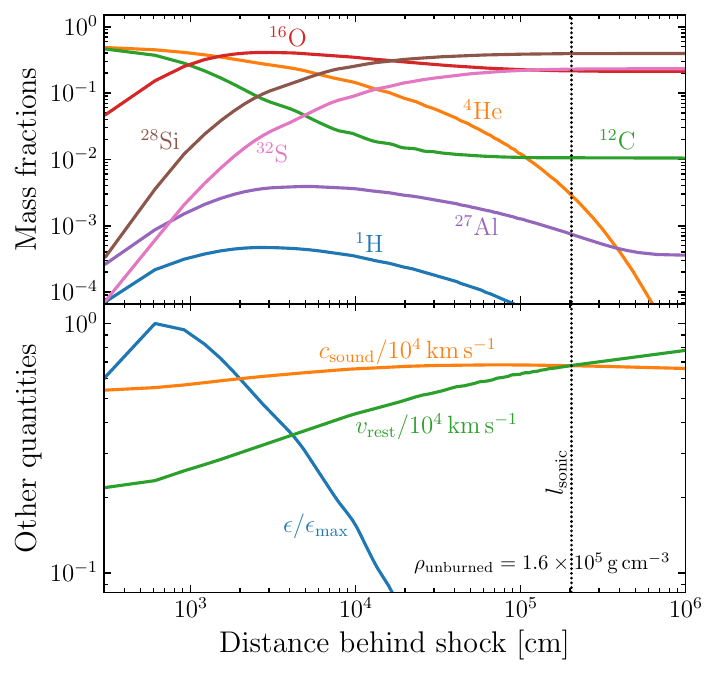}
	\caption{Profiles vs.\ distance behind the shock front of a steady-state detonation.  The upstream unburned fuel has equal mass fractions of $^{4}$He and $^{12}$C and a density of $\unit[1.6\E{5}]{g \, cm^{-3}}$.  The top panel shows mass fractions of selected isotopes, and the bottom panel shows the sound speed and rest frame fluid velocity (in units of $\unit[10^4]{km \, s^{-1}}$) and the energy generation rate normalized to the maximum value, $\epsilon/\epsilon_{\rm max}$.  The dotted line shows the location where the rest frame fluid velocity is equal to the sound speed.}
	\label{fig:profiles}
\end{figure}

Figure \ref{fig:profiles} shows the results for our fiducial case with an initial unburned fuel density of $ \unit[1.6\E{5}]{g \, cm^{-3}}$ and a minimum cell size of 300\,cm.  The initial composition is equal mass fractions of $^4$He and $^{12}$C, as with all of the calculations discussed in this section.  The top panel shows the steady-state post-shock profiles of mass fractions for selected isotopes, and the bottom panel shows the energy generation rate, normalized to its maximum value ($\epsilon/\epsilon_{\rm max}$) as well as the sound speed ($c_{\rm sound}$) and the fluid velocity in the rest frame of the shock ($v_{\rm rest}$), both in units of $ \unit[10^4]{km \, s^{-1}}$.

The primary products of this detonation are $^{28}$Si, $^{32}$S, and $^{16}$O, with a smaller amount of $^{24}$Mg, $^{20}$Ne, $^{36}$Ar (not shown in Fig.\ \ref{fig:profiles}) and unburnt $^{12}$C.  Essentially no iron-group elements are produced; the heaviest isotope that reaches a mass fraction $>10^{-3}$ is $^{40}$Ca.  Crucially, a small amount of hydrogen is generated, as  can be seen in the top panel, peaking at a mass fraction of $5\E{-4}$.  These protons are produced via $^{24}{\rm Mg}(\alpha,p)^{27}$Al and, to a lesser extent, $^{20} {\rm Ne} (\alpha,p)^{23}$Na, reactions; the small number of  $^{20}$Ne and $^{24}$Mg seeds are generated via direct $\alpha$-captures starting from $^{12}$C.  We note that while \cite{shen14b} emphasized the importance of $^{14}$N as a source of protons via the $^{14} {\rm N} (\alpha,p) ^{17} {\rm O}$ reaction, our calculations show that free protons can be generated even if no $^{14}$N is available.

As discussed by \cite{wbs06} in the context of Type I X-ray bursts and by \cite{shen14b} for helium shell detonations on WDs, these protons greatly accelerate helium-burning by allowing for the $^{12} {\rm C} (p,\gamma)^{13} {\rm N}(\alpha,p)^{16}$O pathway, shortcutting the relatively slow direct $\alpha$-capture on $^{12}$C.  \cite{shen14b} noted that this can reduce the helium-burning timescale by a factor of $\sim 5000$ at a density of $\unit[10^5]{g \, cm^{-3}}$ when compared to the triple-$\alpha$ reaction (see their Fig.\ 1), but this was for an assumed composition of $X_{\rm 12C} = 0.05$ and $X_{\rm H}=10^{-4}$.  In the transition region between the helium-rich shell and the carbon-rich core, the carbon mass fraction is a factor of 10 higher and the hydrogen mass fraction grows to a factor of $\sim 5$ larger as well, reducing the burning time to just $\sim 3\E{-6}$ of the triple-$\alpha$ timescale.  It is thus vital that any studies of helium detonations, especially those occurring in thin shells, use large nuclear reaction networks; nuclear networks that shortcut $\alpha$-chain burning and  only assume direct $\alpha$-captures and $(\alpha,p)(p,\gamma)$ reactions yield qualitatively different and incorrect results in this regime.

The bottom panel of Figure \ref{fig:profiles} shows that the fluid velocity in the shock's rest frame reaches the local sound speed at a distance behind the shock front that we label the sonic lengthscale, $l_{\rm sonic}$.  In a standard Chapman-Jouguet (C-J) detonation, this lengthscale coincides with  the complete conversion of fuel to ash \citep{fd79}.  In our case, a small amount of burning continues beyond this point (as evidenced by the decreasing $^4$He mass fraction), but this region is out of sonic contact with the shock front, and the energy release does not contribute to its propagation.  This lengthscale is thus closely related to the C-J lengthscale, but we label it $l_{\rm sonic}$ to avoid any confusion.  

\begin{figure}
  \centering
  \includegraphics[width=\columnwidth]{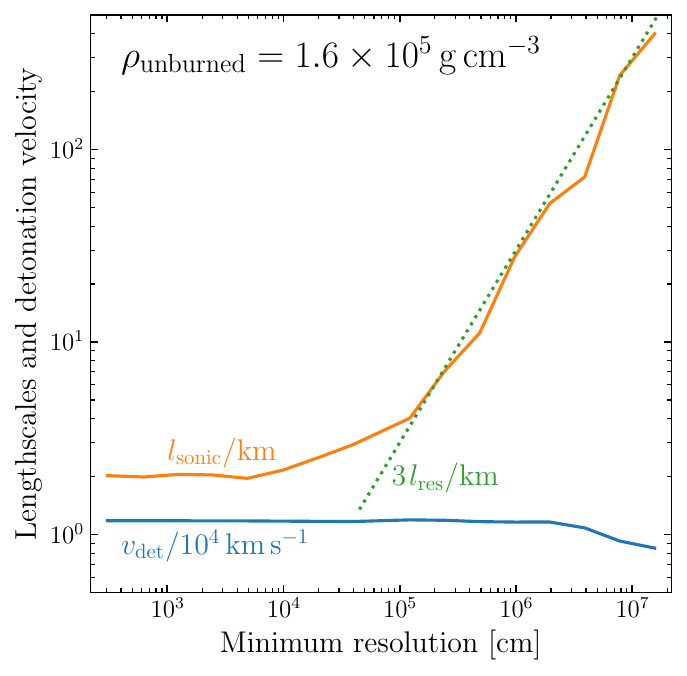}
	\caption{Resolution study for a detonation with the same initial conditions as for the calculation shown in Figure \ref{fig:profiles}.  The detonation velocity in units of $ \unit[10^4]{km \, s^{-1}}$ is shown in blue, and the sonic lengthscale in km is shown in orange.  The green dotted line is three times the minimum cell size in km.}
	\label{fig:restest}
\end{figure}

Figure \ref{fig:restest} shows the results of a resolution study for our fiducial one-dimensional detonation calculation ($\rho_{\rm unburned}= \unit[1.6\E{5}]{g \, cm^{-3}}$, initial $X_{\rm 4He}=X_{\rm 12C}=0.5$).  As the minimum resolution, $l_{\rm res}$, coarsens and the sonic lengthscale is no longer resolved, $l_{\rm sonic}$ becomes $\sim 3 \times l_{\rm res}$.  That is to say, the energy release that powers the detonation occurs over just three zones in unresolved helium detonations.

The total energy release and the commensurate detonation velocity remain close to the actual, physical value even for grossly unresolved simulations: e.g., when the minimum resolution is 10 times larger than the resolved sonic lengthscale, the detonation velocity is still 99\% of the converged result.  Even at a minimum resolution of 100 times the true value of $l_{\rm sonic}$, the velocity is 70\% of the converged value.

We do not find any cases in which a  detonation that successfully propagates at a coarse resolution  fails to propagate at a finer resolution.  Our resolution studies show that, for simulations that do not resolve the true sonic lengthscale, going to higher resolutions increases the detonation velocity and decreases the sonic lengthscale.

Previous studies have probed the success and failure of helium shell detonations, pointing out the importance of post-shock radial expansion as a mechanism for quenching the burning and causing a detonation to fail \citep{tmb12,moor13a,shen14b}.  If radial expansion occurs within sonic communication of the detonation front, it can act as an additional energy sink that weakens and ultimately quenches the propagating detonation.

For a strong shock, as appropriate for the detonations we consider, the jump conditions yield a pressure in the burned ash of $P_{\rm burn} = \rho_{\rm unburned} v_{\rm det}^2 / (\gamma_{\rm burn}+1)$, where  $\gamma_{\rm burn}$ is the equation of state adiabatic exponent, $\gamma = 1 + P/(u \rho)$, evaluated in the ash.  The overpressured ash expands radially with an acceleration of $g P_{\rm burn} / P_{\rm unburned}$, where $g$ is the gravitational acceleration at the interface and $P_{\rm unburned}$ is the unperturbed pressure.  Thus, the ash will expand a pressure scale height ($H_P=P_{\rm unburned}/\rho_{\rm unburned} g$) in a timescale 
\begin{align}
	t_{\rm expand} & \sim \sqrt{ \frac{ 2H_P P_{\rm unburned} }{ P_{\rm burn} g } } \nonumber \\
	& \sim \sqrt{  2 (\gamma_{\rm burn}+1)  }  \frac{H_P}{v_{\rm det}}  ,
\end{align}
and so the lengthscale that the detonation traverses in the time that the overpressure causes significant radial expansion is
\begin{align}
	l_{\rm expand} \sim  \sqrt{  2(\gamma_{\rm burn}+1)  } H_P  .
\end{align}
Thus, for a shell detonation to successfully propagate, it must complete (or nearly complete) its burning before the detonation has moved a lengthscale equivalent to the pressure scale height in the unburned medium: that is, a successful shell detonation requires $l_{\rm sonic} \lesssim H_P$.

\begin{figure}
  \centering
  \includegraphics[width=\columnwidth]{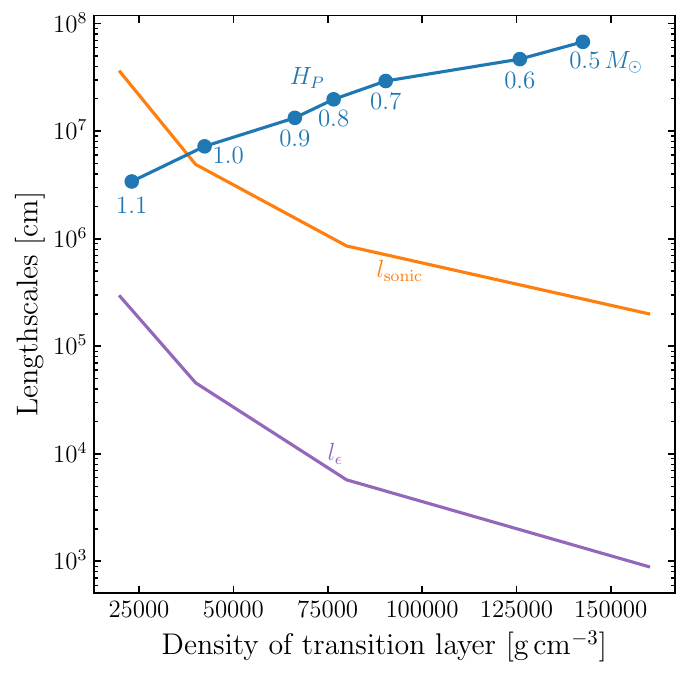}
	\caption{Various lengthscales vs.\ the density of the transition layer for the WD models discussed in Section \ref{sec:mesa}.  The pressure scale height is shown in blue, with each mass as labeled.  The orange line shows the sonic lengthscale assuming equal mass fractions of $^4$He and $^{12}$C, and the purple line shows the lengthscale between the shock and the location of maximal energy generation.}
	\label{fig:lvsrho}
\end{figure}

Figure \ref{fig:lvsrho} shows the detonation lengthscales $l_{\rm sonic}$ and $l_\epsilon$  (the post-shock distance to the location of maximum energy generation rate)  and the scale height at the transition layer for the WD models discussed in Section \ref{sec:mesa} as a function of the density of the unburned fuel in the case of the detonation lengthscales and the density of the transition layers for the WD models.  We see that lower-mass WDs $ \lesssim 1.0 \msol$ are born with helium shells for which the transition regions have relatively large scale heights.  If a detonation can be initiated in such a transition region, it should successfully propagate without being affected by the radial expansion caused by the shock's overpressure.  Conversely, for higher-mass WDs $\gtrsim 1.0 \msol$, the radial blowout will quench the burning, and no detonation is possible unless more mass is added to the shell, which both decreases $l_{\rm sonic}$ and increases the scale height.

% -----------------------------------------------------------
% -----------------------------------------------------------

\section{Two-dimensional detonations in realistic C/O WDs}
\label{sec:2D}

In this section, we test the intuition gained from our one-dimensional toy calculations with two-dimensional axisymmetric simulations performed with \flash.   We use the same 56-isotope nuclear network as in our one-dimensional calculations.  We do not use tracer particles \citep{shen18a,boos21a,boos24a}, as we do not attempt to track detailed nucleosynthesis in this work.  The composition, density, and temperature profiles calculated with \mesa, described in Section \ref{sec:mesa}, are mapped onto grids with domain sizes $\sim 1.2$ times larger than the WD radii.  The shell detonations are initiated by perturbing stellar material within hotspots centered on the symmetry axis at the transition layer.  The hotspots have radii of $4\E{7}$\,cm, maximal temperatures at the center of $2\E{9}$\,K, linear temperature gradients, and densities equal to four times the local density.  These hotspot conditions are extreme, but the present work focuses only on the success or failure of the shell detonation's propagation and the shock's convergence within the core.  We leave a proper examination of the initiation of the detonation to other complementary studies.

% -----------------------------------------------------------
% -----------------------------------------------------------

\subsection{Simulations of models with natal profiles}

\begin{figure}
  \centering
  \includegraphics[width=\columnwidth]{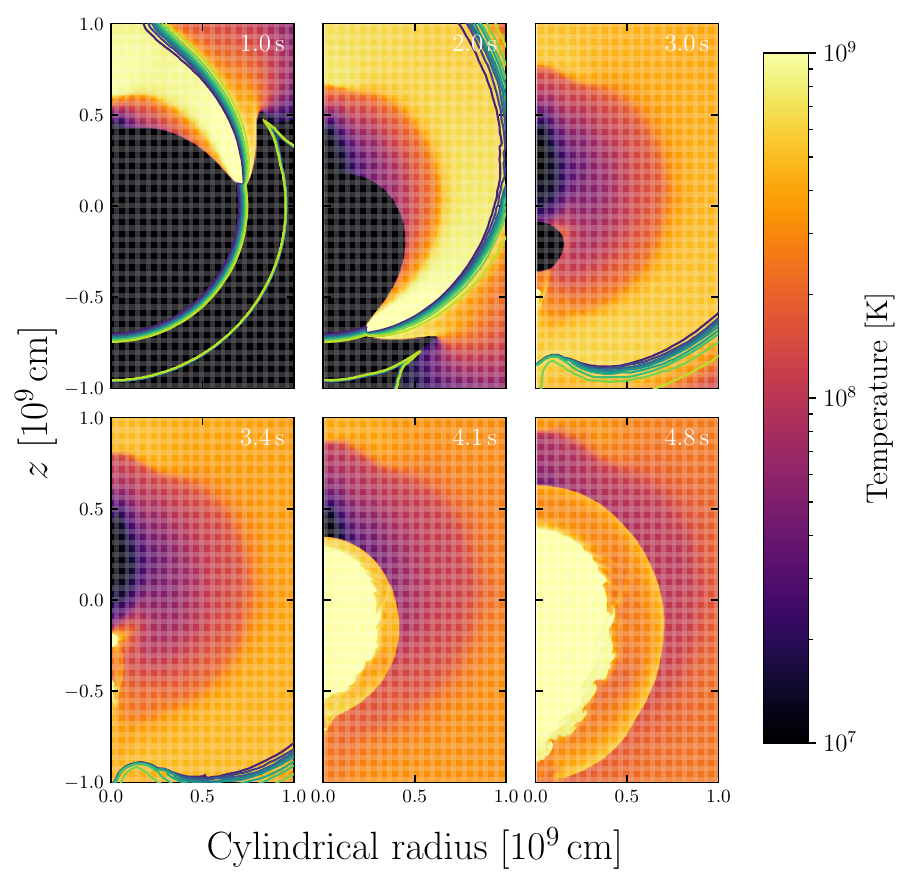}
	\caption{Temperatures (colors) and helium mass fractions (green contours) for six snapshots of a $0.5 \msol$ double detonation simulation.  Helium mass fraction contours increase linearly from a value of $0.1$ to $0.9$ from dark to light.}
	\label{fig:multid}
\end{figure}

Figure \ref{fig:multid} shows multiple snapshots of the temperature for our $0.5 \msol$ model with a minimum cell size of 23\,km.  The nine contours show linearly increasing constant helium mass fractions of $0.1-0.9$ from dark to light colors.  In the first two panels (1.0 and 2.0\,s after the simulation begins), the shell detonation can be seen to propagate successfully, led by a front in the transition layer where $^4$He and $^{12}$C have  roughly equal mass fractions. Our goal is not to calculate detailed nucleosynthesis in these simulations, but we note that the shell detonation primarily produces $^{16}$O, $^{28}$Si, and $^{32}$S with essentially no iron-group elements, matching the expectations from our one-dimensional calculations.  By 3.0\,s, the shell detonation has completed its journey and the shock wave it sent into the core focuses towards a convergence region.  The converging shock ignites a core detonation just before 3.4\,s.  In the final two panels (4.1 and 4.8\,s), the core detonation has extinguished due to the relatively low densities in this low-mass model and has become an outwardly propagating shock wave, as evidenced by the $> \unit[10^9]{K}$ central ash surrounded by an expanding shell of just-shocked, but non-burned, core material at a temperature of $\sim \unit[3\E{8}]{K}$.

\begin{deluxetable*}{cccccccc}
\tablecolumns{8}
\tablecaption{Results of multi-dimensional simulations\label{tab:res}}
\tablehead{
\colhead{Total} & \colhead{Mass above} & \colhead{Transition} & \colhead{Minimum} & \colhead{Shell} & \colhead{Average shell}  & \colhead{$z$-coordinate of}  &\colhead{Core}     \\[-0.25cm]
\colhead{mass} & \colhead{transition layer} & \colhead{layer density} & \colhead{resolution} & \colhead{detonation?}  & \colhead{detonation velocity}  & \colhead{core convergence}  &\colhead{detonation?}     \\[-0.25cm]
\colhead{[$M_\odot$] } & \colhead{[$M_\odot$]} & \colhead{[g\,cm$^{-3}$]} & \colhead{[km]} & & \colhead{[1000\,km\,s$^{-1}$]}  & \colhead{[1000\,km]}  &\colhead{}    }
\startdata
0.5 & $4.138\E{-2}$  & $ 1.426\E{5} $  & $375$ & N & --- & ---  & ---  \\
  &    &    & $188$ & Y & $8.166$ & $-2.69$  & Y (with 47\,km min.\ res.)  \\
  &    &    & $93.8$ &  Y &$8.924$ &  $-2.45$  & Y (with  47\,km min.\ res.)  \\
  &    &    & $46.9$ &  Y &$9.636$ & $-2.32 $ & Y \\
  &    &    & $23.4$ & Y & $10.39$ &  $-2.18$  & Y \\
0.6 & $2.214\E{-2}$  & $ 1.259\E{5} $  & $328$ & N & --- & ---  & ---  \\
  &    &    & $164$ & Y & $8.100$ &  $-2.85$  & Y (with  41\,km min.\ res.)  \\
  &    &    & $82.0$ & Y & $8.816$ &  $-2.61 $ & Y (with   41\,km min.\ res.)  \\
  &    &    & $41.0$ & Y & $9.451$ &  $-2.44 $ & Y \\
  &    &    & $20.5$ & Y & $10.14$ &  $-2.32 $ & Y \\
0.7 & $8.469\E{-3}$  & $ 9.036\E{4} $  & $71.1$ & N & --- & ---  & ---  \\
  &    &    & $35.6$ &  Y &$8.938$  & $-2.56$  & Y (with   18\,km min.\ res.)  \\
  &    &    & $17.8$ &  Y &$9.856$  & $-2.42 $ & Y  \\
0.8 & $4.070\E{-3}$  & $ 7.656\E{4} $  & $64.1$ & N & --- & ---  & ---  \\
  &    &    & $32.0$ &  Y &$8.708$  & $-2.54$  & Y (with   8.0\,km min.\ res.)  \\
  &    &    & $16.0$ &  Y &$9.685$  & $-2.40 $ & Y (with   8.0\,km min.\ res.)  \\
0.9 & $1.900\E{-3}$  & $ 6.629\E{4} $  & $58.6$ & N & --- & ---  & ---  \\
  &    &    & $29.3$ &  Y &$8.531$  & $-2.53$  & Y (with   3.7\,km min.\ res.)  \\
  &    &    & $14.7$ &  Y &$9.408$  & $-2.37 $ & Y (with   3.7\,km min.\ res.)  \\
1.0 & $5.232\E{-4}$  & $ 4.237\E{4} $  & $25.4$ & N & --- & ---  & ---  \\
  &    &    & $12.7$ & N & --- & ---  & ---  \\
  &    &    & $6.35$ & N & --- & ---  & ---  \\
  &    &    & $3.17$ & N & --- & ---  & ---  \\
$1.0+0.001$ & $1.504\E{-3}$  & $ 8.337\E{4} $  & $54.7$ & N & --- & ---  & ---  \\
  &    &    & $27.3$ & Y & $7.580$  & $-2.54$  & Y (with   3.4\,km min.\ res.)  \\
  &    &    & $13.7$ & Y & $8.333$  & $-2.40$  & Y (with   3.4\,km min.\ res.)  \\
1.1 & $9.412\E{-5}$  & $ 2.307\E{4} $  & 46.9  & N  & ---  & ---   & ---   \\
  &    &    & 23.4  & N  & ---  & ---   & ---   \\
  &    &    & 11.7  & N  & ---  & ---   & ---   \\
$1.1+0.001$ & $1.099\E{-3}$  & $ 1.031\E{5} $  & $39.1$ & N & --- & ---  & ---  \\
  &    &    & $19.5$ & N & --- & ---  & ---  \\
  &    &    & $9.77$ & N & --- & ---  & ---  \\
  &    &    & $4.88$ & N & --- & ---  & ---  \\
$1.1+0.002$ & $ 2.110\E{-3}$  & $ 1.582\E{5} $  & $23.4$ & N & --- & ---  & ---  \\
  &    &    & $11.7$ &  Y &$7.789$  & $-2.25$  & Y (with   1.5\,km min.\ res.)  \\
$1.1+0.003$ & $  3.126\E{-3}$  & $  2.065\E{5} $  & $93.8$ & N & --- & ---  & ---  \\
  &    &    & $46.9$ & Y & $7.624$  & $-2.29$  & Y (with   3.0\,km min.\ res.)  \\
  &    &    & $23.4$ &  Y &$8.123$  & $-2.21 $ & Y (with   3.0\,km min.\ res.)  \\
  &    &    & $11.7$ & Y & $8.676$  & $-2.13  $ & Y (with   3.0\,km min.\ res.)  \\
\enddata
\end{deluxetable*}

\begin{figure}
  \centering
  \includegraphics[width=\columnwidth]{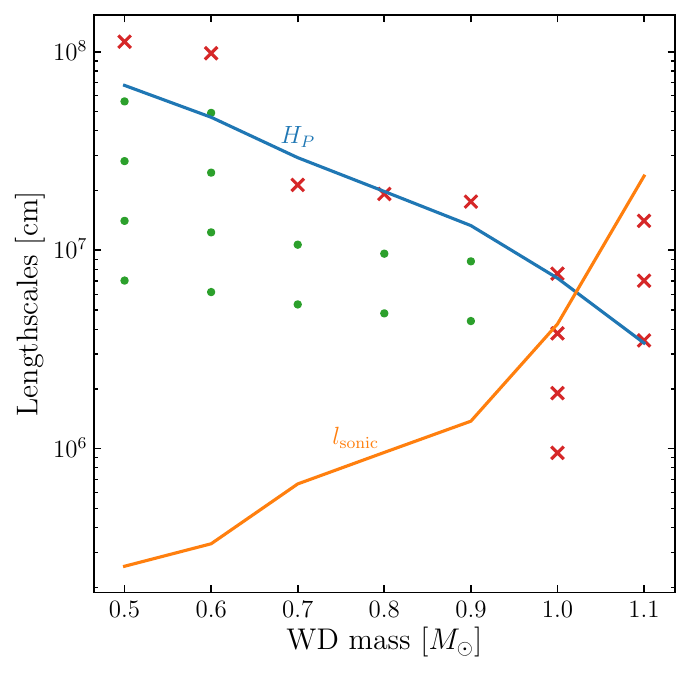}
	\caption{Various lengthscales vs.\ total WD mass.  The blue line shows the pressure scale height measured at the transition layer.  The sonic lengthscale of a detonation powered by helium/carbon fuel at the transition layer density is shown in orange.  Crosses and circles show three times the minimum cell size for two-dimensional simulations with failed and successful detonations, respectively.}
	\label{fig:lvsm}
\end{figure}

Table~\ref{tab:res} and Figure~\ref{fig:lvsm} summarize the results of our two-dimensional simulations.  In Figure~\ref{fig:lvsm}, the pressure scale height at the transition layer for each WD model is shown in blue, and the orange line shows the sonic lengthscale for a detonation at the transition layer density.  Crosses and circles show three times the minimum cell size --  approximately the sonic lengthscale for unresolved detonations -- for failed and successful simulations, respectively.

As expected from our previous discussion, we find that lower-mass models $ \lesssim 1.0 \msol$ can support successful shell detonations with their natal compositional profiles, given sufficient resolution such that $l_{\rm sonic} \lesssim H_P$, where $l_{\rm sonic} \sim 3 \times l_{\rm res}$ for unresolved detonations.  Conversely, higher-mass models $ \gtrsim 1.0 \msol$ cannot have successful shell detonations at birth because the scale height at the transition layer is shorter than the sonic lengthscale of a detonation at that density, and the post-shock radial expansion extinguishes the burning.    Table~\ref{tab:res} also demonstrates that, for  cases where the shell detonates, the converging shock always ignites a core detonation.  For some models, achieving the core detonation requires decreasing the minimum cell size prior to shock convergence (see, e.g., \citealt{boos21a} and \citealt{riva22a}), as denoted in the table, but, for a sufficient resolution, a shell detonation implies a core detonation \citep{shen14a,ghos22a}.   We note that in some previous studies \citep{pakm13a,pakm21a,pakm22a,roy22a}, a successful shell detonation failed to trigger a converging core detonation.  However, the minimum spatial resolutions in these studies range from $10-30$\,km, whereas Table~\ref{tab:res} shows that achieving core detonations  can require resolutions of $\sim 3$\, km.  Thus, the lack of core detonations in these previous studies is likely due to insufficient resolution, as speculated by Pakmor et al.

We also note that, for simulations with successful detonations, higher resolutions imply more complete shell burning and faster detonation velocities (column 6 of Table \ref{tab:res}).  This is in line with the resolution study described in Section \ref{sec:1D} and Figure \ref{fig:restest}.  The shorter time for the shell detonation to completely encircle the WD core also implies a convergence point closer to the WD's center (column 7 of Table \ref{tab:res}), which will have implications for the ejecta structure of the resulting explosion, affecting, for example, the asymmetries in spectral velocities from opposing lines of sight.

We find successful detonations for significantly smaller shell masses than the minimum detonatable masses presented by \cite{shen14b}.  Our $0.9 \msol$ model, with $M_{\rm He,tot}=0.0026 \msol$ and $M_{\rm He,transition}=0.0019 \msol$, successfully detonates while \cite{shen14b} find a minimum critical shell mass of $0.007 \msol$ for this core mass.  This is due to the different composition of the detonating layer: \cite{shen14b} assumed a composition of $X_{\rm 4He} = 0.891$, $X_{\rm 12C} = X_{\rm 16O} = 0.05$, and $X_{\rm 14N} = 0.009$, whereas in our realistic WD models, the detonation propagates within a layer of nearly equal $^4$He and $^{12}$C mass fractions.  Even without $^{14}$N, sufficient protons are still produced to allow the $^{12}{\rm C}(p,\gamma)(\alpha,p)^{16}{\rm O}$ to occur, and the higher initial mass fraction of $^{12}$C allows for more complete consumption of the $^4$He at lower densities, thus powering a successful detonation.

The 0.5 and $0.6 \msol$ core detonations produce $1.2\E{-4}$ and $1.3\E{-3} \msol$ of $^{56}$Ni, respectively.   Although our simulations are not designed to calculate detailed nucleosynthesis,\footnote{Our simulations are not followed all the way to homologous expansion, nor do we include tracer particles for post-processing with a large reaction network.  We refer the interested reader to previous work focused on nucleosynthesis of exploding sub-Chandrasekhar-mass WDs, such as \cite{shen18a}, \cite{gron21b}, \cite{boos21a,boos24a}, and \cite{keeg23a}.} our findings of very small $^{56}$Ni masses for low-mass core detonations are likely robust, given the low densities $< \unit[10^7]{g \, cm^{-3}}$ \citep{seit17a}.  However, these yields will be boosted if the low-mass core is the companion WD in a double WD binary in which the primary explodes.  The impact of the first explosion's ejecta may both initiate a shell detonation of the secondary and send a strong shock into the core that increases its density prior to the ignition of the core detonation.  This explains why \cite{shen18a} find a $^{56}$Ni mass of $0.02 \msol$ for their $0.8 \msol$ WD detonation whereas \cite{boos24a}'s $0.8 \msol$ secondary WD, ignited after the impact of a primary WD's ejecta, produces $0.15 \msol$ of $^{56}$Ni.

% -----------------------------------------------------------
% -----------------------------------------------------------

\subsection{High-mass models with added mass}

We supplement the $1.0$ and $1.1 \msol$ calculations with additional models in which we add a small amount of helium-rich mass ($X_{\rm 4He}=0.99$, $X_{\rm 14N}=0.01$) and then allow the resulting stars to relax back to their initial central temperatures with mixing and diffusion active.  Such stars may arise in binaries where the companion is a helium-burning star that transfers material when it expands and crosses the helium-burning equivalent of the Hertzsprung gap \citep{yoon03a,ruit13a}.  The first phase of this mass transfer may result in stable helium-burning, growing the WD accretor by $\sim 0.1 \msol$ \citep{pier14a,broo16a}.  However, as mass transfer continues, the accretion rate declines, stable helium-burning ends, and unstable helium novae will result.\footnote{We assume that these novae do not result in the merger of the helium star and WD, whereas this may be the case for novae in double WD systems due to orbital angular momentum loss during the nova \citep{shen15a,nele16a,schr16a,shen22a}.}  Eventually, mass transfer ceases, and the resulting helium layer on the WD's surface will be, on average, several $10^{-3} \msol$ larger than the minimum mass for stable helium-burning \citep{it89,bild07} that the WDs initially possessed at birth.

\begin{figure}
  \centering
  \includegraphics[width=\columnwidth]{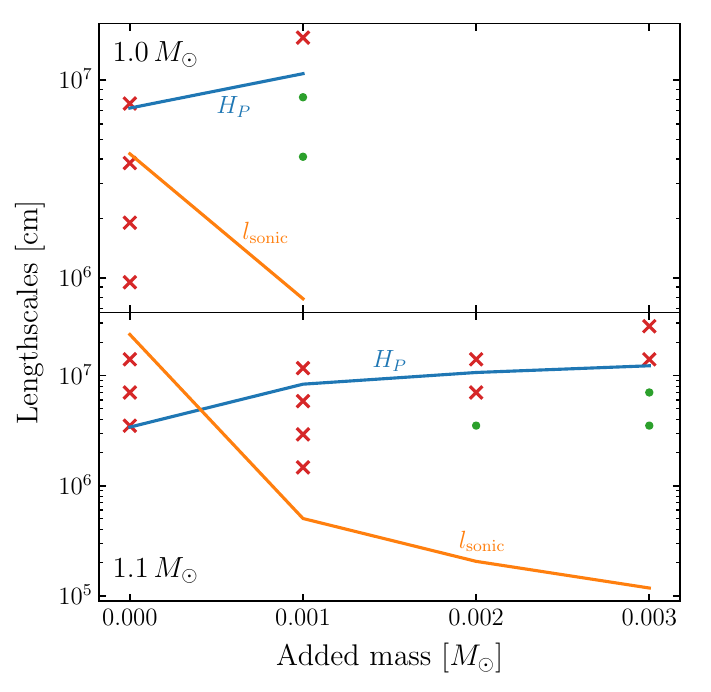}
	\caption{Lengthscales vs.\ added mass for $1.0$ (top panel) and $1.1 \msol$ (bottom panel) WDs.  The pressure scale heights at the transition radius are shown as blue lines, and the orange lines demarcate the sonic lengthscale at the transition radius's density.  Crosses and circles represent three times the minimum cell size for simulations with failed and successful detonations, respectively.}
	\label{fig:lvsdm}
\end{figure}

The additional mass increases the density of the transition layer, which increases the scale height and decreases the detonation's sonic lengthscale, both of which help to make shell detonations more successful.  Table~\ref{tab:res} and Figure \ref{fig:lvsdm} summarize the results of these additional simulations.  Adding $10^{-3} \msol$ to the $1.0 \msol$ model allows for a successful shell detonation (and subsequent core detonation).  For reasons that are unclear, an additional $ \geq 2\E{-3} \msol$ is required for a shell and core detonation for the $1.1 \msol$ model, even though $l_{\rm sonic} < H_P$ is achieved for the $1.1 + 10^{-3} \msol$ case.  Further investigation is required to understand why a detonation does not propagate in this case.

% -----------------------------------------------------------
% -----------------------------------------------------------

\section{Discussion}
\label{sec:disc}

Our results show that most C/O WDs are born with structures that allow for successful double detonations if a helium shell detonation can be initiated.  Only massive C/O WDs $\gtrsim 1.0 \msol$ do not possess sufficiently dense transition regions for successful shell detonations at birth.  However, some of these massive WDs will accrete enough helium after they are first formed and prior to the formation of the double WD binary to allow for successful shell and core detonations during the merging process.

Thus, the crucial issue determining  the  success of a double detonation and subsequent SN~Ia is reduced to: does the unstable mass transfer during  a double WD merger reach conditions to initiate a shell detonation?  Explorations of detonation ignition during double WD mergers have been performed \citep{guil10,dan12,pakm13a,shen14b,dan15a,iwat22a,pakm22a,roy22a} and are still ongoing (Rajavel et al.\ submitted).  Further work needs to be carried out, including the possible ignition of a subsonic deflagration prior to the onset of the detonation, but some broad trends can be inferred.  For one, a more massive accretor will yield a hotter and denser region surrounding the impacting accretion stream.  This is due to the fact that a more massive WD has a deeper gravitational potential, so the virial temperature and the ram pressure, and thus the depth and density the stream reaches, will be higher.  Additionally, binaries with higher mass ratios yield accretion streams that impact the accretor's surface more perpendicularly  and thus plunge more deeply.

Thus, the hottest and densest hotspots with the greatest likelihood to ignite detonations  will be generated in binaries with the highest-mass accretors and donors.  This may be the reason that there do not appear to be more subluminous SNe~Ia \citep{li11b,ghos22a,shar22a}, which likely occur in binaries with primary WD masses $\sim 0.85 \msol$ \citep{blon17a,shen18a,shen21a,shen21b}, which should outnumber binaries with more massive WDs.   Even though we have found that double detonations can propagate in all C/O WDs $\lesssim 1.0 \msol$, the ignition of such a detonation in a low-mass WD via an accretion stream may require a mass ratio close to unity or may even be entirely impossible for a low-enough primary mass (although see \citealt{mora24a} for a counter-example).

If the primary WD does explode, our results suggest that if the impacting ejecta can ignite a helium detonation in the companion's helium shell (which seems plausible given the outer ejecta velocities $> \unit[10^4]{km \, s^{-1}}$), the detonation will successfully propagate in the companion's natal transition layer and  lead to a double detonation of the companion, resulting in a two-star explosion \citep{tani19a,pakm22a,boos24a}.  If this occurs most of the time, and thus most SNe~Ia leave no companion behind, then historical SN remnants will contain no surviving companion WDs, which appears to be the case for the remnants of SN~1006 \citep{kerz18a,shie22a} and SNR 0509-67.5 \citep{shie23a}.  Exceptions to this outcome are if the donor is a low-mass helium core WD (for which densities are relatively low and  $^{12}$C is not present to accelerate helium-burning\footnote{Several studies \citep{papi15a,tani19a,boos24a} have suggested the possibility that the primary WD's impacting ejecta can indeed ignite  a detonation in a helium-core WD companion.  However, in all three studies, ignition was only achieved at artificially reduced separations between the two WDs.  Further study is necessary to examine this possibility, which perhaps exists for the highest-mass helium-core WDs but likely does not for lower-mass cases.}) or a high-mass C/O WD $\gtrsim 1.0 \msol$.  One other exception is if the primary WD is $\gtrsim 1.0 \msol$ and did not undergo an additional phase of helium accretion prior to the merger.  In this case, a direct carbon ignition might be triggered \citep{pakm10,pakm11,pakm12b} but would require a significant amount of mass transfer from the companion, possibly removing the helium-rich layers and thus preventing a double detonation when the primary's ejecta impacts the companion (if the companion has not been completely tidally disrupted at the time of explosion).  The interaction of the ejecta with the excess  C/O-rich companion material surrounding the exploding primary WD will increase the SN luminosity  and possibly form dust at late times \citep{taub13b,sieb24a} and may be the explanation for the rare class of ``super-Chandrasekhar-mass'' SN~2003fg-like SNe~Ia \citep{howe06a,taub13b,rask14a,noeb16a,fitz23a}.

These three cases of exceptions may explain the class of seven recently discovered hypervelocity SN survivors \citep{shen18b,baue21a,chan22a,elba23a,wern24a}.  The relatively low-velocity ($\unit[1050]{km \, s^{-1}}$), possibly helium-rich \citep{chan22a} hypervelocity star D6-2 \citep{shen18b} may have been a  helium-core WD donor.  Meanwhile, the four  higher-velocity ($\sim \unit[2000]{km \, s^{-1}}$) hot ($> \unit[2\E{4}]{K}$) candidates discovered by \cite{elba23a} may have been high-mass donors with insufficient helium layers to undergo a double detonation upon impact.  Finally, D6-1 and D6-3 \citep{shen18b}, which are both cooler like D6-2 ($< \unit[10^4]{K}$) but have much higher velocities ($\sim \unit[2000]{km \, s^{-1}}$), may be formerly high-mass companions that lost significant mass while triggering direct carbon ignition of their primaries, leading to 2003fg-like SNe~Ia;   indeed, the late-time spectra of the SN~2003fg-like SN~2020hvf have been interpreted as evidence for a wind from a  surviving star \citep{shen17a,sieb23a}.  Further studies are necessary to explore all of these possible outcomes.

% -----------------------------------------------------------
% -----------------------------------------------------------

\section{Conclusions}
\label{sec:conc}

In this paper, we have calculated realistic models of C/O WDs for a range of   masses and then conducted one- and two-dimensional detonation simulations based on these results.  We have found that:

\begin{itemize}

\item{Most  C/O WDs are born with composition and density profiles that can support a detonation in the transition layer between the helium-rich shell and carbon-rich core.  If such a shell detonation can be triggered, we find that it will, in turn, trigger a core detonation.  No additional accretion after the formation of the WD is necessary for these detonations to propagate.}

\item{Large nuclear reaction networks and realistic compositional profiles are crucial for capturing the propagation of these relatively low-density detonations in hydrodynamic codes.}

\item{Only higher-mass C/O WDs $\gtrsim 1.0 \msol$ cannot support shell detonations at birth.  However, $\sim 10^{-3} \msol$ of material accreted prior to the formation of the double WD binary, which is expected for some systems, including all of those with primary C/O WDs $>1.05 \msol$, is enough to allow for successful shell detonations when the double WDs begin to merge.}

\item{If a companion C/O WD $ \lesssim 1.0 \msol$ is impacted with sufficient strength by the primary WD's ejecta and has not transferred substantial mass prior to the explosion, it will undergo its own double detonation, leading to a two-star SN~Ia.}

\item{The detonations of natal low-mass helium shells do not produce significant amounts of iron-group elements.  The detonations of low-mass C/O cores also do not produce a significant amount of $^{56}$Ni, but the yield will be increased if the core detonation is preceded by a strong shock from the impact of a primary WD explosion.}

\end{itemize}

Combined with the extensive existing literature comparing observations to predicted observables of thin-shell sub-Chandrasekhar-mass WD explosions \citep{fhr07,fink10,krom10,sim10,blon17a,shen18a,poli19a,town19a,gron20a,gron21a,shen21a,shen21b,boos21a,pakm22a,boos24a}, our work suggests that a majority of SNe~Ia,  ranging from subluminous SN~1991bg-likes to overluminous SN~1991T-likes and SN~2003fg-likes, may arise from the double detonations of both WDs in a merging double WD binary.  A smaller fraction of SNe~Ia may result from the explosion of only the more massive WD, leaving behind surviving hypervelocity companion WDs.

Our present study has several limitations, including the two-dimensional nature of our explosion simulations and our assumption that a detonation can be initiated in the shell-core transition layer by mass transfer from a companion WD.  The interaction of the accretion stream with the accretor's surface layers is a three-dimensional, turbulent, and complex process, which we do not attempt to capture in this paper.  The propagation of the detonations we model may be influenced by these fluctuations, including compositional inhomogeneities mixing the core and surface material.  Deflagrations or failed detonations may also alter the structure of the transition layer prior to the onset of a successful detonation.  Future work should include an examination of detonation ignition conditions during mass transfer and during the interaction of the primary WD's ejecta  with  the companion; two-dimensional and three-dimensional explosion modeling and radiative transfer  of two-star explosions, including non-local thermodynamic equilibrium and early- and late-time calculations; simulations of primary WD explosions within transferred companion material, possibly appropriate for 2003fg-like ``super-Chandrasekhar-mass'' SNe~Ia; and stellar evolution calculations of hypervelocity survivors.  These and other studies remain to be done, but it appears hopeful that the resolution to the SN~Ia progenitor question is close at hand.

% -----------------------------------------------------------
% -----------------------------------------------------------

\software{\texttt{matplotlib} \citep{hunt07a}, \mesa \citep{paxt11,paxt13,paxt15a,paxt18a,paxt19a}, \texttt{FLASH} \citep{fryx00,dube09a}}

\begin{acknowledgments}

We thank Alison Miller for helpful discussions.  This work was supported by NASA through the Astrophysics Theory Program (80NSSC20K0544)  and by NASA/ESA Hubble Space Telescope programs \#15871 and \#15918.  D.M.T.\ also acknowledges support from the National Science Foundation under Grant No.\ 2307442.   This research used the Savio computational cluster resource provided by the Berkeley Research Computing program at the University of California, Berkeley (supported by the UC Berkeley Chancellor, Vice Chancellor of Research, and Office of the CIO).

\end{acknowledgments}

% -----------------------------------------------------------
% -----------------------------------------------------------

% -----------------------------------------------------------
% -----------------------------------------------------------


\begin{thebibliography}{}
\expandafter\ifx\csname natexlab\endcsname\relax\def\natexlab#1{#1}\fi

\bibitem[{{Bauer} {et~al.}(2021){Bauer}, {Chandra}, {Shen}, \&
  {Hermes}}]{baue21a}
{Bauer}, E.~B., {Chandra}, V., {Shen}, K.~J., \& {Hermes}, J.~J. 2021, \apjl,
  923, L34

\bibitem[{{Bildsten} {et~al.}(2007){Bildsten}, {Shen}, {Weinberg}, \&
  {Nelemans}}]{bild07}
{Bildsten}, L., {Shen}, K.~J., {Weinberg}, N.~N., \& {Nelemans}, G. 2007,
  \apjl, 662, L95

\bibitem[{{Blondin} {et~al.}(2017){Blondin}, {Dessart}, {Hillier}, \&
  {Khokhlov}}]{blon17a}
{Blondin}, S., {Dessart}, L., {Hillier}, D.~J., \& {Khokhlov}, A.~M. 2017,
  \mnras, 470, 157

\bibitem[{{Boos} {et~al.}(2024){Boos}, {Townsley}, \& {Shen}}]{boos24a}
{Boos}, S.~J., {Townsley}, D.~M., \& {Shen}, K.~J. 2024, submitted
  (arXiv:2401.08011)

\bibitem[{{Boos} {et~al.}(2021){Boos}, {Townsley}, {Shen}, {Caldwell}, \&
  {Miles}}]{boos21a}
{Boos}, S.~J., {Townsley}, D.~M., {Shen}, K.~J., {Caldwell}, S., \& {Miles},
  B.~J. 2021, \apj, 919, 126

\bibitem[{{Brooks} {et~al.}(2016){Brooks}, {Bildsten}, {Schwab}, \&
  {Paxton}}]{broo16a}
{Brooks}, J., {Bildsten}, L., {Schwab}, J., \& {Paxton}, B. 2016, \apj, 821, 28

\bibitem[{{Burmester} {et~al.}(2023){Burmester}, {Ferrario}, {Pakmor},
  {Seitenzahl}, {Ruiter}, \& {Hole}}]{burm23a}
{Burmester}, U.~P., {Ferrario}, L., {Pakmor}, R., {et~al.} 2023, \mnras, 523,
  527

\bibitem[{{Chandra} {et~al.}(2022){Chandra}, {Hwang}, {Zakamska}, {Blouin},
  {Swan}, {Marsh}, {Shen}, {G{\"a}nsicke}, {Hermes}, {Putterman}, {Bauer},
  {Petrosky}, {Dhillon}, {Littlefair}, \& {Ashley}}]{chan22a}
{Chandra}, V., {Hwang}, H.-C., {Zakamska}, N.~L., {et~al.} 2022, \mnras, 512,
  6122

\bibitem[{{Chugunov} {et~al.}(2007){Chugunov}, {Dewitt}, \&
  {Yakovlev}}]{chug07a}
{Chugunov}, A.~I., {Dewitt}, H.~E., \& {Yakovlev}, D.~G. 2007, \prd, 76, 025028

\bibitem[{{Dan} {et~al.}(2015){Dan}, {Guillochon}, {Br{\"u}ggen},
  {Ramirez-Ruiz}, \& {Rosswog}}]{dan15a}
{Dan}, M., {Guillochon}, J., {Br{\"u}ggen}, M., {Ramirez-Ruiz}, E., \&
  {Rosswog}, S. 2015, \mnras, 454, 4411

\bibitem[{{Dan} {et~al.}(2011){Dan}, {Rosswog}, {Guillochon}, \&
  {Ramirez-Ruiz}}]{dan11}
{Dan}, M., {Rosswog}, S., {Guillochon}, J., \& {Ramirez-Ruiz}, E. 2011, \apj,
  737, 89

\bibitem[{{Dan} {et~al.}(2012){Dan}, {Rosswog}, {Guillochon}, \&
  {Ramirez-Ruiz}}]{dan12}
---. 2012, \mnras, 422, 2417

\bibitem[{{Doherty} {et~al.}(2015){Doherty}, {Gil-Pons}, {Siess}, {Lattanzio},
  \& {Lau}}]{dohe15a}
{Doherty}, C.~L., {Gil-Pons}, P., {Siess}, L., {Lattanzio}, J.~C., \& {Lau}, H.
  H.~B. 2015, \mnras, 446, 2599

\bibitem[{Dubey {et~al.}(2009)Dubey, Reid, Weide, Antypas, Ganapathy, Riley,
  Sheeler, \& Siegal}]{dube09a}
Dubey, A., Reid, L.~B., Weide, K., {et~al.} 2009, Parallel Computing, 35, 512

\bibitem[{{El-Badry} {et~al.}(2023){El-Badry}, {Shen}, {Chandra}, {Bauer},
  {Fuller}, {Strader}, {Chomiuk}, {Naidu}, {Caiazzo}, {Rodriguez}, {Nagarajan},
  {Yamaguchi}, {Vanderbosch}, {Roulston}, {G{\"a}nsicke}, {Han}, {Burdge},
  {Filippenko}, {Brink}, \& {Zheng}}]{elba23a}
{El-Badry}, K., {Shen}, K.~J., {Chandra}, V., {et~al.} 2023, The Open Journal
  of Astrophysics, 6, 28

\bibitem[{{Fickett} \& {Davis}(1979)}]{fd79}
{Fickett}, W., \& {Davis}, C. 1979, Detonation, Los Alamos Series in Basic and
  Applied Sciences (Berkeley, CA: Univ. California)

\bibitem[{{Fink} {et~al.}(2007){Fink}, {Hillebrandt}, \& {R{\"o}pke}}]{fhr07}
{Fink}, M., {Hillebrandt}, W., \& {R{\"o}pke}, F.~K. 2007, \aap, 476, 1133

\bibitem[{{Fink} {et~al.}(2010){Fink}, {R{\"o}pke}, {Hillebrandt},
  {Seitenzahl}, {Sim}, \& {Kromer}}]{fink10}
{Fink}, M., {R{\"o}pke}, F.~K., {Hillebrandt}, W., {et~al.} 2010, \aap, 514,
  A53

\bibitem[{{Fitz Axen} \& {Nugent}(2023)}]{fitz23a}
{Fitz Axen}, M., \& {Nugent}, P. 2023, \apj, 953, 13

\bibitem[{{Fryxell} {et~al.}(2000){Fryxell}, {Olson}, {Ricker}, {Timmes},
  {Zingale}, {Lamb}, {MacNeice}, {Rosner}, {Truran}, \& {Tufo}}]{fryx00}
{Fryxell}, B., {Olson}, K., {Ricker}, P., {et~al.} 2000, \apjs, 131, 273

\bibitem[{{Ghosh} \& {Kushnir}(2022)}]{ghos22a}
{Ghosh}, A., \& {Kushnir}, D. 2022, \mnras, 515, 286

\bibitem[{{Gronow} {et~al.}(2020){Gronow}, {Collins}, {Ohlmann}, {Pakmor},
  {Kromer}, {Seitenzahl}, {Sim}, \& {R{\"o}pke}}]{gron20a}
{Gronow}, S., {Collins}, C., {Ohlmann}, S.~T., {et~al.} 2020, \aap, 635, A169

\bibitem[{{Gronow} {et~al.}(2021{\natexlab{a}}){Gronow}, {Collins}, {Sim}, \&
  {Roepke}}]{gron21a}
{Gronow}, S., {Collins}, C.~E., {Sim}, S.~A., \& {Roepke}, F.~K.
  2021{\natexlab{a}}, \aap, submitted (arXiv:2102.06719), arXiv:2102.06719

\bibitem[{{Gronow} {et~al.}(2021{\natexlab{b}}){Gronow}, {Cote}, {Lach},
  {Seitenzahl}, {Collins}, {Sim}, \& {Roepke}}]{gron21b}
{Gronow}, S., {Cote}, B., {Lach}, F., {et~al.} 2021{\natexlab{b}}, \aap,
  submitted (arXiv:2103.14050), arXiv:2103.14050

\bibitem[{{Guillochon} {et~al.}(2010){Guillochon}, {Dan}, {Ramirez-Ruiz}, \&
  {Rosswog}}]{guil10}
{Guillochon}, J., {Dan}, M., {Ramirez-Ruiz}, E., \& {Rosswog}, S. 2010, \apjl,
  709, L64

\bibitem[{{H\"{o}flich} \& {Khokhlov}(1996)}]{hk96}
{H\"{o}flich}, P., \& {Khokhlov}, A. 1996, \apj, 457, 500

\bibitem[{{Howell} {et~al.}(2006){Howell}, {Sullivan}, {Nugent}, {Ellis},
  {Conley}, {Le Borgne}, {Carlberg}, {Guy}, {Balam}, {Basa}, {Fouchez}, {Hook},
  {Hsiao}, {Neill}, {Pain}, {Perrett}, \& {Pritchet}}]{howe06a}
{Howell}, D.~A., {Sullivan}, M., {Nugent}, P.~E., {et~al.} 2006, \nat, 443, 308

\bibitem[{Hunter(2007)}]{hunt07a}
Hunter, J.~D. 2007, Computing in Science \& Engineering, 9, 90

\bibitem[{{Iben} \& {Tutukov}(1984)}]{it84}
{Iben}, Jr., I., \& {Tutukov}, A.~V. 1984, \apjs, 54, 335

\bibitem[{{Iben} \& {Tutukov}(1989)}]{it89}
---. 1989, \apj, 342, 430

\bibitem[{{Iwata} \& {Maeda}(2022)}]{iwat22a}
{Iwata}, K., \& {Maeda}, K. 2022, \apj, 941, 87

\bibitem[{{Keegans} {et~al.}(2023){Keegans}, {Pignatari}, {Stancliffe},
  {Travaglio}, {Jones}, {Gibson}, {Townsley}, {Miles}, {Shen}, \&
  {Few}}]{keeg23a}
{Keegans}, J.~D., {Pignatari}, M., {Stancliffe}, R.~J., {et~al.} 2023, \apjs,
  268, 8

\bibitem[{{Kerzendorf} {et~al.}(2014){Kerzendorf}, {Childress},
  {Scharw{\"a}chter}, {Do}, \& {Schmidt}}]{kerz14c}
{Kerzendorf}, W.~E., {Childress}, M., {Scharw{\"a}chter}, J., {Do}, T., \&
  {Schmidt}, B.~P. 2014, \apj, 782, 27

\bibitem[{{Kerzendorf} {et~al.}(2012){Kerzendorf}, {Schmidt}, {Laird},
  {Podsiadlowski}, \& {Bessell}}]{kerz12a}
{Kerzendorf}, W.~E., {Schmidt}, B.~P., {Laird}, J.~B., {Podsiadlowski}, P., \&
  {Bessell}, M.~S. 2012, \apj, 759, 7

\bibitem[{{Kerzendorf} {et~al.}(2018){Kerzendorf}, {Strampelli}, {Shen},
  {Schwab}, {Pakmor}, {Do}, {Buchner}, \& {Rest}}]{kerz18a}
{Kerzendorf}, W.~E., {Strampelli}, G., {Shen}, K.~J., {et~al.} 2018, \mnras,
  479, 192

\bibitem[{{Kerzendorf} {et~al.}(2013){Kerzendorf}, {Yong}, {Schmidt}, {Simon},
  {Jeffery}, {Anderson}, {Podsiadlowski}, {Gal-Yam}, {Silverman}, {Filippenko},
  {Nomoto}, {Murphy}, {Bessell}, {Venn}, \& {Foley}}]{kerz13a}
{Kerzendorf}, W.~E., {Yong}, D., {Schmidt}, B.~P., {et~al.} 2013, \apj, 774, 99

\bibitem[{{Kippenhahn} {et~al.}(2013){Kippenhahn}, {Weigert}, \&
  {Weiss}}]{kipp13a}
{Kippenhahn}, R., {Weigert}, A., \& {Weiss}, A. 2013, {Stellar Structure and
  Evolution} (Berlin: Springer), doi:10.1007/978-3-642-30304-3

\bibitem[{{Kromer} {et~al.}(2010){Kromer}, {Sim}, {Fink}, {R{\"o}pke},
  {Seitenzahl}, \& {Hillebrandt}}]{krom10}
{Kromer}, M., {Sim}, S.~A., {Fink}, M., {et~al.} 2010, \apj, 719, 1067

\bibitem[{{Kushnir} \& {Katz}(2020)}]{kush20b}
{Kushnir}, D., \& {Katz}, B. 2020, \mnras, 493, 5413

\bibitem[{{Kushnir} {et~al.}(2013){Kushnir}, {Katz}, {Dong}, {Livne}, \&
  {Fern{\'a}ndez}}]{kush13a}
{Kushnir}, D., {Katz}, B., {Dong}, S., {Livne}, E., \& {Fern{\'a}ndez}, R.
  2013, \apjl, 778, L37

\bibitem[{{Li} {et~al.}(2011){Li}, {Leaman}, {Chornock}, {Filippenko},
  {Poznanski}, {Ganeshalingam}, {Wang}, {Modjaz}, {Jha}, {Foley}, \&
  {Smith}}]{li11b}
{Li}, W., {Leaman}, J., {Chornock}, R., {et~al.} 2011, \mnras, 412, 1441

\bibitem[{{Liu} {et~al.}(2023){Liu}, {R{\"o}pke}, \& {Han}}]{liu23b}
{Liu}, Z.-W., {R{\"o}pke}, F.~K., \& {Han}, Z. 2023, Research in Astronomy and
  Astrophysics, 23, 082001

\bibitem[{{Livne}(1990)}]{livn90}
{Livne}, E. 1990, \apjl, 354, L53

\bibitem[{{Moll} {et~al.}(2014){Moll}, {Raskin}, {Kasen}, \&
  {Woosley}}]{moll14a}
{Moll}, R., {Raskin}, C., {Kasen}, D., \& {Woosley}, S.~E. 2014, \apj, 785, 105

\bibitem[{{Moore} {et~al.}(2013){Moore}, {Townsley}, \& {Bildsten}}]{moor13a}
{Moore}, K., {Townsley}, D.~M., \& {Bildsten}, L. 2013, \apj, 776, 97

\bibitem[{{Mor{\'a}n-Fraile} {et~al.}(2024){Mor{\'a}n-Fraile}, {Holas},
  {R{\"o}pke}, {Pakmor}, \& {Schneider}}]{mora24a}
{Mor{\'a}n-Fraile}, J., {Holas}, A., {R{\"o}pke}, F.~K., {Pakmor}, R., \&
  {Schneider}, F. R.~N. 2024, \aap, 683, A44

\bibitem[{{Murai} {et~al.}(1968){Murai}, {Sugimoto}, {H{\={o}}shi}, \&
  {Hayashi}}]{mura68a}
{Murai}, T., {Sugimoto}, D., {H{\={o}}shi}, R., \& {Hayashi}, C. 1968, Progress
  of Theoretical Physics, 39, 619

\bibitem[{{Nelemans} {et~al.}(2016){Nelemans}, {Siess}, {Repetto}, {Toonen}, \&
  {Phinney}}]{nele16a}
{Nelemans}, G., {Siess}, L., {Repetto}, S., {Toonen}, S., \& {Phinney}, E.~S.
  2016, \apj, 817, 69

\bibitem[{{Noebauer} {et~al.}(2016){Noebauer}, {Taubenberger}, {Blinnikov},
  {Sorokina}, \& {Hillebrandt}}]{noeb16a}
{Noebauer}, U.~M., {Taubenberger}, S., {Blinnikov}, S., {Sorokina}, E., \&
  {Hillebrandt}, W. 2016, \mnras, 463, 2972

\bibitem[{{Nomoto}(1982)}]{nomo82b}
{Nomoto}, K. 1982, \apj, 257, 780

\bibitem[{{Nomoto} {et~al.}(1984){Nomoto}, {Thielemann}, \& {Yokoi}}]{nty84}
{Nomoto}, K., {Thielemann}, F.-K., \& {Yokoi}, K. 1984, \apj, 286, 644

\bibitem[{{Nugent} {et~al.}(1997){Nugent}, {Baron}, {Branch}, {Fisher}, \&
  {Hauschildt}}]{nuge97}
{Nugent}, P., {Baron}, E., {Branch}, D., {Fisher}, A., \& {Hauschildt}, P.~H.
  1997, \apj, 485, 812

\bibitem[{{Pakmor} {et~al.}(2011){Pakmor}, {Hachinger}, {R{\"o}pke}, \&
  {Hillebrandt}}]{pakm11}
{Pakmor}, R., {Hachinger}, S., {R{\"o}pke}, F.~K., \& {Hillebrandt}, W. 2011,
  \aap, 528, A117

\bibitem[{{Pakmor} {et~al.}(2010){Pakmor}, {Kromer}, {R{\"o}pke}, {Sim},
  {Ruiter}, \& {Hillebrandt}}]{pakm10}
{Pakmor}, R., {Kromer}, M., {R{\"o}pke}, F.~K., {et~al.} 2010, \nat, 463, 61

\bibitem[{{Pakmor} {et~al.}(2012){Pakmor}, {Kromer}, {Taubenberger}, {Sim},
  {R{\"o}pke}, \& {Hillebrandt}}]{pakm12b}
{Pakmor}, R., {Kromer}, M., {Taubenberger}, S., {et~al.} 2012, \apjl, 747, L10

\bibitem[{{Pakmor} {et~al.}(2013){Pakmor}, {Kromer}, {Taubenberger}, \&
  {Springel}}]{pakm13a}
{Pakmor}, R., {Kromer}, M., {Taubenberger}, S., \& {Springel}, V. 2013, \apjl,
  770, L8

\bibitem[{{Pakmor} {et~al.}(2021){Pakmor}, {Zenati}, {Perets}, \&
  {Toonen}}]{pakm21a}
{Pakmor}, R., {Zenati}, Y., {Perets}, H.~B., \& {Toonen}, S. 2021, \mnras, 503,
  4734

\bibitem[{{Pakmor} {et~al.}(2022){Pakmor}, {Callan}, {Collins}, {de Mink},
  {Holas}, {Kerzendorf}, {Kromer}, {Neunteufel}, {O'Brien}, {R{\"o}pke},
  {Ruiter}, {Seitenzahl}, {Shingles}, {Sim}, \& {Taubenberger}}]{pakm22a}
{Pakmor}, R., {Callan}, F.~P., {Collins}, C.~E., {et~al.} 2022, \mnras, 517,
  5260

\bibitem[{{Papish} {et~al.}(2015){Papish}, {Soker}, {Garc{\'{\i}}a-Berro}, \&
  {Aznar-Sigu{\'a}n}}]{papi15a}
{Papish}, O., {Soker}, N., {Garc{\'{\i}}a-Berro}, E., \& {Aznar-Sigu{\'a}n}, G.
  2015, \mnras, 449, 942

\bibitem[{{Paxton} {et~al.}(2011){Paxton}, {Bildsten}, {Dotter}, {Herwig},
  {Lesaffre}, \& {Timmes}}]{paxt11}
{Paxton}, B., {Bildsten}, L., {Dotter}, A., {et~al.} 2011, \apjs, 192, 3

\bibitem[{{Paxton} {et~al.}(2013){Paxton}, {Cantiello}, {Arras}, {Bildsten},
  {Brown}, {Dotter}, {Mankovich}, {Montgomery}, {Stello}, {Timmes}, \&
  {Townsend}}]{paxt13}
{Paxton}, B., {Cantiello}, M., {Arras}, P., {et~al.} 2013, \apjs, 208, 4

\bibitem[{{Paxton} {et~al.}(2015){Paxton}, {Marchant}, {Schwab}, {Bauer},
  {Bildsten}, {Cantiello}, {Dessart}, {Farmer}, {Hu}, {Langer}, {Townsend},
  {Townsley}, \& {Timmes}}]{paxt15a}
{Paxton}, B., {Marchant}, P., {Schwab}, J., {et~al.} 2015, \apjs, 220, 15

\bibitem[{{Paxton} {et~al.}(2018){Paxton}, {Schwab}, {Bauer}, {Bildsten},
  {Blinnikov}, {Duffell}, {Farmer}, {Goldberg}, {Marchant}, {Sorokina},
  {Thoul}, {Townsend}, \& {Timmes}}]{paxt18a}
{Paxton}, B., {Schwab}, J., {Bauer}, E.~B., {et~al.} 2018, \apjs, 234, 34

\bibitem[{{Paxton} {et~al.}(2019){Paxton}, {Smolec}, {Schwab}, {Gautschy},
  {Bildsten}, {Cantiello}, {Dotter}, {Farmer}, {Goldberg}, {Jermyn}, {Kanbur},
  {Marchant}, {Thoul}, {Townsend}, {Wolf}, {Zhang}, \& {Timmes}}]{paxt19a}
{Paxton}, B., {Smolec}, R., {Schwab}, J., {et~al.} 2019, \apjs, 243, 10

\bibitem[{{Piersanti} {et~al.}(2014){Piersanti}, {Tornamb{\'e}}, \&
  {Yungelson}}]{pier14a}
{Piersanti}, L., {Tornamb{\'e}}, A., \& {Yungelson}, L.~R. 2014, \mnras, 445,
  3239

\bibitem[{{Polin} {et~al.}(2019){Polin}, {Nugent}, \& {Kasen}}]{poli19a}
{Polin}, A., {Nugent}, P., \& {Kasen}, D. 2019, \apj, 873, 84

\bibitem[{{Raskin} {et~al.}(2014){Raskin}, {Kasen}, {Moll}, {Schwab}, \&
  {Woosley}}]{rask14a}
{Raskin}, C., {Kasen}, D., {Moll}, R., {Schwab}, J., \& {Woosley}, S. 2014,
  \apj, 788, 75

\bibitem[{{Raskin} {et~al.}(2012){Raskin}, {Scannapieco}, {Fryer},
  {Rockefeller}, \& {Timmes}}]{rask12}
{Raskin}, C., {Scannapieco}, E., {Fryer}, C., {Rockefeller}, G., \& {Timmes},
  F.~X. 2012, \apj, 746, 62

\bibitem[{{Rivas} {et~al.}(2022){Rivas}, {Harris}, {Hix}, \&
  {Messer}}]{riva22a}
{Rivas}, F., {Harris}, J.~A., {Hix}, W.~R., \& {Messer}, O.~E.~B. 2022, \apj,
  937, 2

\bibitem[{{Roy} {et~al.}(2022){Roy}, {Tiwari}, {Bobrick}, {Kosakowski},
  {Fisher}, {Perets}, {Kashyap}, {Lor{\'e}n-Aguilar}, \&
  {Garc{\'\i}a-Berro}}]{roy22a}
{Roy}, N.~C., {Tiwari}, V., {Bobrick}, A., {et~al.} 2022, \apjl, 932, L24

\bibitem[{{Ruiter} {et~al.}(2013){Ruiter}, {Sim}, {Pakmor}, {Kromer},
  {Seitenzahl}, {Belczynski}, {Fink}, {Herzog}, {Hillebrandt}, {R{\"o}pke}, \&
  {Taubenberger}}]{ruit13a}
{Ruiter}, A.~J., {Sim}, S.~A., {Pakmor}, R., {et~al.} 2013, \mnras, 429, 1425

\bibitem[{{Schaefer} \& {Pagnotta}(2012)}]{sp12}
{Schaefer}, B.~E., \& {Pagnotta}, A. 2012, \nat, 481, 164

\bibitem[{{Schreiber} {et~al.}(2016){Schreiber}, {Zorotovic}, \&
  {Wijnen}}]{schr16a}
{Schreiber}, M.~R., {Zorotovic}, M., \& {Wijnen}, T.~P.~G. 2016, \mnras, 455,
  L16

\bibitem[{{Seitenzahl} \& {Townsley}(2017)}]{seit17a}
{Seitenzahl}, I.~R., \& {Townsley}, D.~M. 2017, {in Handbook of Supernovae, ed.
  A.~W. Alsabti \& P. Murdin} (New York: Springer), 1955

\bibitem[{{Sharon} \& {Kushnir}(2022)}]{shar22a}
{Sharon}, A., \& {Kushnir}, D. 2022, \mnras, 509, 5275

\bibitem[{{Shen}(2015)}]{shen15a}
{Shen}, K.~J. 2015, \apjl, 805, L6

\bibitem[{{Shen} \& {Bildsten}(2007)}]{sb07}
{Shen}, K.~J., \& {Bildsten}, L. 2007, \apj, 660, 1444

\bibitem[{{Shen} \& {Bildsten}(2009)}]{sb09b}
---. 2009, \apj, 699, 1365

\bibitem[{{Shen} \& {Bildsten}(2014)}]{shen14a}
---. 2014, \apj, 785, 61

\bibitem[{{Shen} {et~al.}(2021{\natexlab{a}}){Shen}, {Blondin}, {Kasen},
  {Dessart}, {Townsley}, {Boos}, \& {Hillier}}]{shen21a}
{Shen}, K.~J., {Blondin}, S., {Kasen}, D., {et~al.} 2021{\natexlab{a}}, \apjl,
  909, L18

\bibitem[{{Shen} {et~al.}(2023){Shen}, {Blouin}, \& {Breivik}}]{shen23a}
{Shen}, K.~J., {Blouin}, S., \& {Breivik}, K. 2023, \apjl, 955, L33

\bibitem[{{Shen} {et~al.}(2021{\natexlab{b}}){Shen}, {Boos}, {Townsley}, \&
  {Kasen}}]{shen21b}
{Shen}, K.~J., {Boos}, S.~J., {Townsley}, D.~M., \& {Kasen}, D.
  2021{\natexlab{b}}, \apj, 922, 68

\bibitem[{{Shen} {et~al.}(2018{\natexlab{a}}){Shen}, {Kasen}, {Miles}, \&
  {Townsley}}]{shen18a}
{Shen}, K.~J., {Kasen}, D., {Miles}, B.~J., \& {Townsley}, D.~M.
  2018{\natexlab{a}}, \apj, 854, 52

\bibitem[{{Shen} {et~al.}(2010){Shen}, {Kasen}, {Weinberg}, {Bildsten}, \&
  {Scannapieco}}]{shen10}
{Shen}, K.~J., {Kasen}, D., {Weinberg}, N.~N., {Bildsten}, L., \&
  {Scannapieco}, E. 2010, \apj, 715, 767

\bibitem[{{Shen} \& {Moore}(2014)}]{shen14b}
{Shen}, K.~J., \& {Moore}, K. 2014, \apj, 797, 46

\bibitem[{{Shen} \& {Quataert}(2022)}]{shen22a}
{Shen}, K.~J., \& {Quataert}, E. 2022, \apj, 938, 31

\bibitem[{{Shen} \& {Schwab}(2017)}]{shen17a}
{Shen}, K.~J., \& {Schwab}, J. 2017, \apj, 834, 180

\bibitem[{{Shen} {et~al.}(2018{\natexlab{b}}){Shen}, {Boubert}, {G{\"a}nsicke},
  {Jha}, {Andrews}, {Chomiuk}, {Foley}, {Fraser}, {Gromadzki}, {Guillochon},
  {Kotze}, {Maguire}, {Siebert}, {Smith}, {Strader}, {Badenes}, {Kerzendorf},
  {Koester}, {Kromer}, {Miles}, {Pakmor}, {Schwab}, {Toloza}, {Toonen},
  {Townsley}, \& {Williams}}]{shen18b}
{Shen}, K.~J., {Boubert}, D., {G{\"a}nsicke}, B.~T., {et~al.}
  2018{\natexlab{b}}, \apj, 865, 15

\bibitem[{{Shields} {et~al.}(2023){Shields}, {Arunachalam}, {Kerzendorf},
  {Hughes}, {Biriouk}, {Monk}, \& {Buchner}}]{shie23a}
{Shields}, J.~V., {Arunachalam}, P., {Kerzendorf}, W., {et~al.} 2023, \apjl,
  950, L10

\bibitem[{{Shields} {et~al.}(2022){Shields}, {Kerzendorf}, {Hosek}, {Shen},
  {Rest}, {Do}, {Lu}, {Fullard}, {Strampelli}, \& {Zenteno}}]{shie22a}
{Shields}, J.~V., {Kerzendorf}, W., {Hosek}, M.~W., {et~al.} 2022, \apjl, 933,
  L31

\bibitem[{{Siebert} {et~al.}(2023){Siebert}, {Foley}, {Zenati}, {Dimitriadis},
  {Schmidt}, {Yang}, {Davis}, {Taggart}, \& {Rojas-Bravo}}]{sieb23a}
{Siebert}, M.~R., {Foley}, R.~J., {Zenati}, Y., {et~al.} 2023, \apj, 958, 173

\bibitem[{{Siebert} {et~al.}(2024){Siebert}, {Kwok}, {Johansson}, {Jha},
  {Blondin}, {Dessart}, {Foley}, {Hillier}, {Larison}, {Pakmor}, {Temim},
  {Andrews}, {Auchettl}, {Badenes}, {Barna}, {Bostroem}, {Brenner Newman},
  {Brink}, {Bustamante-Rosell}, {Camacho-Neves}, {Clocchiatti}, {Coulter},
  {Davis}, {Deckers}, {Dimitriadis}, {Dong}, {Farah}, {Filippenko},
  {Fl{\"o}rs}, {Fox}, {Garnavich}, {Padilla Gonzalez}, {Graur}, {Hambsch},
  {Hosseinzadeh}, {Howell}, {Hughes}, {Kerzendorf}, {Le Saux}, {Maeda},
  {Maguire}, {McCully}, {Mihalenko}, {Newsome}, {O'Brien}, {Pearson},
  {Pellegrino}, {Pierel}, {Polin}, {Rest}, {Rojas-Bravo}, {Sand}, {Schwab},
  {Shahbandeh}, {Shrestha}, {Smith}, {Strolger}, {Szalai}, {Taggart},
  {Terreran}, {Terwel}, {Tinyanont}, {Valenti}, {Vink{\'o}}, {Wheeler}, {Yang},
  {Zheng}, {Ashall}, {DerKacy}, {Galbany}, {Hoeflich}, {Hsiao}, {de Jaeger},
  {Lu}, {Maund}, {Medler}, {Morrell}, {Shappee}, {Stritzinger}, {Suntzeff},
  {Tucker}, \& {Wang}}]{sieb24a}
{Siebert}, M.~R., {Kwok}, L.~A., {Johansson}, J., {et~al.} 2024, \apj, 960, 88

\bibitem[{{Sim} {et~al.}(2010){Sim}, {R{\"o}pke}, {Hillebrandt}, {Kromer},
  {Pakmor}, {Fink}, {Ruiter}, \& {Seitenzahl}}]{sim10}
{Sim}, S.~A., {R{\"o}pke}, F.~K., {Hillebrandt}, W., {et~al.} 2010, \apjl, 714,
  L52

\bibitem[{{Tanikawa} {et~al.}(2018){Tanikawa}, {Nomoto}, \&
  {Nakasato}}]{tani18b}
{Tanikawa}, A., {Nomoto}, K., \& {Nakasato}, N. 2018, \apj, 868, 90

\bibitem[{{Tanikawa} {et~al.}(2019){Tanikawa}, {Nomoto}, {Nakasato}, \&
  {Maeda}}]{tani19a}
{Tanikawa}, A., {Nomoto}, K., {Nakasato}, N., \& {Maeda}, K. 2019, \apj, 885,
  103

\bibitem[{{Taubenberger}(2017)}]{taub17a}
{Taubenberger}, S. 2017, {in Handbook of Supernovae, ed. A.~W. Alsabti \& P.
  Murdin} (New York: Springer), 317

\bibitem[{{Taubenberger} {et~al.}(2013){Taubenberger}, {Kromer}, {Hachinger},
  {Mazzali}, {Benetti}, {Nugent}, {Scalzo}, {Pakmor}, {Stanishev},
  {Spyromilio}, {Bufano}, {Sim}, {Leibundgut}, \& {Hillebrandt}}]{taub13b}
{Taubenberger}, S., {Kromer}, M., {Hachinger}, S., {et~al.} 2013, \mnras, 432,
  3117

\bibitem[{{Townsley} {et~al.}(2019){Townsley}, {Miles}, {Shen}, \&
  {Kasen}}]{town19a}
{Townsley}, D.~M., {Miles}, B.~J., {Shen}, K.~J., \& {Kasen}, D. 2019, \apjl,
  878, L38

\bibitem[{{Townsley} {et~al.}(2012){Townsley}, {Moore}, \& {Bildsten}}]{tmb12}
{Townsley}, D.~M., {Moore}, K., \& {Bildsten}, L. 2012, \apj, 755, 4

\bibitem[{{Webbink}(1984)}]{webb84}
{Webbink}, R.~F. 1984, \apj, 277, 355

\bibitem[{{Weinberg} {et~al.}(2006){Weinberg}, {Bildsten}, \& {Schatz}}]{wbs06}
{Weinberg}, N.~N., {Bildsten}, L., \& {Schatz}, H. 2006, \apj, 639, 1018

\bibitem[{{Werner} {et~al.}(2024){Werner}, {Reindl}, {Rauch}, {El-Badry}, \&
  {B{\'e}dard}}]{wern24a}
{Werner}, K., {Reindl}, N., {Rauch}, T., {El-Badry}, K., \& {B{\'e}dard}, A.
  2024, \aap, 682, A42

\bibitem[{{Whelan} \& {Iben}(1973)}]{wi73}
{Whelan}, J., \& {Iben}, I.~J. 1973, \apj, 186, 1007

\bibitem[{{Woosley} {et~al.}(1986){Woosley}, {Taam}, \& {Weaver}}]{wtw86}
{Woosley}, S.~E., {Taam}, R.~E., \& {Weaver}, T.~A. 1986, \apj, 301, 601

\bibitem[{{Wu} {et~al.}(2022){Wu}, {Xiong}, \& {Wang}}]{wu22a}
{Wu}, C., {Xiong}, H., \& {Wang}, X. 2022, \mnras, 512, 2972

\bibitem[{{Yoon} \& {Langer}(2003)}]{yoon03a}
{Yoon}, S.-C., \& {Langer}, N. 2003, \aap, 412, L53

\bibitem[{{Zenati} {et~al.}(2019){Zenati}, {Toonen}, \& {Perets}}]{zena19b}
{Zenati}, Y., {Toonen}, S., \& {Perets}, H.~B. 2019, \mnras, 482, 1135

\end{thebibliography}
\end{document}